\documentclass[fleqn,10pt]{wlscirep}

\usepackage{latexsym,amsmath}
\usepackage{color}
\usepackage{hyperref}

\newcommand{\av}[1]{\langle #1 \rangle}

\newcommand{\km}{q_\textrm{max}}

\newcommand{\AV}{\av{q^2}/\av{q}}
\newcommand{\LM}{\Lambda_1}

\title{Distinct types of eigenvector localization in networks}

\author[1*]{Romualdo Pastor-Satorras}
\author[2,3]{Claudio Castellano}  

\affil[1]{Departament de F\'{\i}sica, Universitat
  Polit\`ecnica de Catalunya, Campus Nord B4, 08034 Barcelona, Spain}

\affil[2]{Istituto dei Sistemi Complessi (ISC-CNR), Via dei Taurini 19,
  I-00185 Roma, Italy}

\affil[3]{Dipartimento di Fisica, ``Sapienza'' Universit\`a di Roma,
  P.le A. Moro 2, I-00185 Roma, Italy}

\affil[*]{romualdo.pastor@upc.edu}

\begin{abstract}
  The spectral properties of the adjacency matrix provide a trove of
  information about the structure and function of complex networks.  In
  particular, the largest eigenvalue and its associated principal
  eigenvector are crucial in the understanding of nodes centrality and
  the unfolding of dynamical processes.  Here we show that two distinct
  types of localization of the principal eigenvector may occur in
  heterogeneous networks.  For synthetic networks with degree
  distribution $P(q) \sim q^{-\gamma}$, localization occurs on the
  largest hub if $\gamma>5/2$; for $\gamma<5/2$ a new type of
  localization arises on a mesoscopic subgraph associated with the shell
  with the largest index in the $K$-core decomposition.  Similar
  evidence for the existence of distinct localization modes is found in
  the analysis of real-world networks. Our results open a new
  perspective on dynamical processes on networks and on a recently
  proposed alternative measure of node centrality based on the non-backtracking
  matrix.
\end{abstract}

\begin{document}

\flushbottom
\maketitle

\thispagestyle{empty}

\section*{Introduction}

An issue of paramount significance regarding the analysis of networked
systems is the identification of the most important (or
\textit{central}) vertices~\cite{Newman2010}.  The \textit{centrality}
of a vertex may stem from the number of different vertices that can be
reached from it, from the role it plays in the communication between
different parts of the network, or from how closely knit its
neighborhood is.  Following these approaches, different centrality
measures have been defined and exploited, such as degree centrality,
betweenness centrality~\cite{freeman77}, or the $K$-core index and
associated $K$-core decomposition~\cite{Seidman1983269}.  Among those
definitions, one of the most relevant is based on the intuitive notion
that nodes are central when they are connected to other central
nodes. This concept is mathematically encoded in the \textit{eigenvector
  centrality}~\cite{Bonacich72} (EC) of node $i$, defined as the
component $f_i$ of the principal eigenvector (PEV) $\mathbf{f}$
associated with the largest eigenvalue $\Lambda_1$ of the adjacency matrix
$A_{ij}$.  EC is the simplest of a family of centralities based on the
spectral properties of the adjacency matrix including, among others,
Katz's centrality\cite{katz53} and PageRank~\cite{brin98anatomy}.

Apart from providing relevant information about the network
structure~\cite{Newman2010}, the PEV and associated largest eigenvalue
play a fundamental role in the theoretical understanding of the behavior
of dynamical processes, such as
synchronization~\cite{PhysRevE.71.036151} and
spreading~\cite{1284681,Castellano2010}, mediated by complex topologies.
Considerable effort has thus been devoted in recent years to the study
of the spectral properties of heterogeneous
networks~\cite{Farkas01,Chung03,Dorogovtsev03,Kuhn08}.  In this
framework, Goltsev \textit{et al.}~\cite{Goltsev12} (see
also~\cite{Nadakuditi13, Martin14}) have considered the
\textit{localization} of the PEV, i.e., whether its normalization weight
is concentrated on a small subset of nodes or not. More in detail, let
us consider an ensemble of networks of size $N$,
with a PEV $f_i$
normalized as a standard Euclidean vector, i.e. $\sum_i f_i^2 = 1$.
An eigenvector is {\em localized} on a subset $V$
of size $N_V$
if a finite fraction of the normalization weight is concentrated on $V$
($\sum_{i \in V} f_i^2 \sim \mathcal{O}(1)$)
despite the fact that $V$
is not extensive, i.e., $N_V$
is not proportional to $N$.
This includes the case of localization on a finite set of nodes
(i.e. $N_V$
independent of $N$,
$N_V=1$
in the extreme case of localization on a single node), but also the case
of localization on a mesoscopic subset of nodes for which
$N_V \sim N^{\beta}$
with $\beta<1$.
Otherwise, the eigenvector is instead {\em delocalized}, and a finite
fraction of the nodes $N_V \sim N$
contribute to the normalization weight, implying that their components
are $f_i \sim N^{-1/2}$.

In this context, Goltsev \textit{et al.}  \cite{Goltsev12} study the
localization in power-law distributed networks, with a degree
distribution scaling as $P(q) \sim q^{-\gamma}$, for which the leading
eigenvalue $\Lambda_1$ is essentially given by the maximum between $\AV$
and $\sqrt{\km}$, where $\km$ is the largest degree in the
network~\cite{Castellano2010,Chung03}.  For $\gamma>5/2$, where
$\Lambda_1 \sim \sqrt{\km}$, Goltsev \textit{et al.} \cite{Goltsev12}
find that the PEV becomes localized around the hub with degree
$\km$~\cite{Goltsev12}.  On the other hand, they argue that, for
$\gamma<5/2$, when $\Lambda_1 \sim \AV$, the PEV is delocalized. These
observations are relevant in different contexts. Firstly, they point out
a weakness of EC as a measure of centrality for heterogeneous power-law
networks ($\gamma>5/2$), because of the exceedingly large role of the
largest hub~\cite{Martin14}.  On the other hand, in the so-called
quenched mean-field approach \cite{Castellano2010,Ferreira12} to epidemic
spreading on networks, the density of infected individuals in the steady
state can be related to the properties of the PEV~\cite{Goltsev12}.  The
localization occurring for large $\gamma$ implies that the density of
infected individuals in the steady state in those processes might not be
an extensive quantity, casting doubts on the validity of this
theoretical approach and on the actual onset of the endemic infected
state.

Here we show that the localization properties of the adjacency matrix
PEV for heterogeneous (power-law distributed) networks are described by a
picture much more complex than previously believed. In fact, we provide
strong numerical evidence that the EC in heterogeneous networks never
achieves full delocalization. In the case of uncorrelated synthetic
networks with a power-law degree distribution, we obtain, by means of a
finite-size scaling analysis, that for mild levels of heterogeneity
(with $\gamma>5/2$), the EC is strongly localized on the hubs, as
previously argued. For high heterogeneity ($\gamma<5/2$), however, we
point out that the EC, as measured by the components of the PEV, is
highly correlated with the corresponding node's degree. This strong
correlation results in an effective localization on a mesoscopic
subgraph, that can be identified as the shell with the largest index in
the $K$-core decomposition of the network~\cite{Seidman1983269}. 
The paper of Goltsev \textit{et al.} \cite{Goltsev12} is perfectly correct
for what concerns the case $\gamma > 5/2$ but, by only considering the
possibility of localization on a finite set of nodes, could not detect the 
mesoscopic localization occurring for $\gamma < 5/2$.
In order to overcome the localization effects intrinsic of the EC, a new
centrality measure, based on the largest eigenvalue of the Hashimoto, or
non-backtracking, matrix, has been recently proposed~\cite{Martin14}. We
observe that this new centrality is not completely free from
localization effects. Thus, while it almost coincides with the EC for
$\gamma<5/2$, and for $\gamma>5/2$ it avoids the extreme localization
around the hubs shown by the EC, it is still localized in this case in
some mesoscopic subset of nodes, whose characterization calls for
further research. The extension of our analysis to the case of real
world networks is hampered by the fact that usually only one network
instance is available, which prevents performing a finite-size scaling
study. Nevertheless, we numerically argue that also for real networks a
twofold scenario holds, in which the PEV is either localized on the
hubs, or effectively localized on the maximum $K$-core of the network.

\section*{Results}

\subsection*{Eigenvector localization and the inverse participation
  ratio}

A full characterization of an undirected network of size $N$ is given by
its adjacency matrix~\cite{Newman2010} $\mathbf{A}$, whose elements take
the value $A_{ij} = 1$ if nodes $i$ and $j$ are connected by an edge,
and value $A_{ij} = 0$ otherwise. The spectral properties of the
adjacency matrix are defined by the set of eigenvalues $\Lambda_i$, and
associated eigenvectors $\mathbf{f}(\Lambda_i)$, $i = 1, \ldots, N$,
defined by
\begin{equation}
  \mathbf{A} \, \mathbf{f}(\Lambda) = \Lambda \mathbf{f}(\Lambda).
\end{equation}
Since the adjacency matrix is symmetric all
its eigenvalues are real. The largest
of those eigenvalues $\Lambda_1$, is associated with the principal
eigenvector (PEV) which we denote simply by $\mathbf{f}$.

The concept of the localization of the PEV $\mathbf{f}$ translates in
determining whether the value of its normalized components is evenly
distributed among all nodes in the network, or either it attains a large
value on some subset, and is much smaller in all the rest.  While this
concept is quite easy to grasp, assessing it in a single network
instance is a delicate issue because any quantitative definition
involves some degree of arbitrariness.  The task becomes however
straightforward when ensembles of networks of different size can be
generated. In such a case, the localization of the eigenvector
$\mathbf{f}$ associated with the eigenvalue $\Lambda$ can be precisely
assessed by computing the inverse participation ratio (IPR), defined
as~\cite{Goltsev12,Martin14},
\begin{equation}   
  Y_\Lambda = \sum_i f_i^4(\Lambda).   
\label{eq:5} 
\end{equation} 
In the absence of any knowledge about the localization support, it is
possible to determine whether an eigenvector is localized (on some
subset in the network) by studying its inverse participation ratio, as a
function of the system size $N$ and fitting its behavior to a power-law
decay of the form
\begin{equation}   
Y_{\Lambda} (N) \sim N^{-\alpha}.   
\label{eq:Methods_1} 
\end{equation} 
If the eigenvector is delocalized, i.e. for $f_i \sim N^{-1/2}$, the
exponent $\alpha$ is equal to 1.  An exponent $\alpha<1$ is evidence
that some form of localization is taking place.  In the case of extreme
localization on a single node, or on a set of nodes with size $N_V$
independent of the network size $N$, the corresponding components of the
PEV are finite and this implies $Y^\mathrm{loc}_{\Lambda} \sim
\mathcal{O}(1)$, i.e., $\alpha=0$ for $N\to\infty$. Finally, if
localization takes place over a subextensive set of nodes of size $N_V \sim
N^{\beta}$, we expect
\begin{equation}   
Y_{\Lambda} (N) \sim \frac{1}{N_V} \sim N^{-\beta},
\label{eq:Methods_2} 
  \end{equation} 
leading to a decay exponent $\alpha = \beta$.

\subsection*{Eigenvector localization in synthetic networks}

\begin{figure}[t]
  \centering
  \includegraphics[clip=true,trim=0cm 0cm 0cm 0cm,width=15cm]{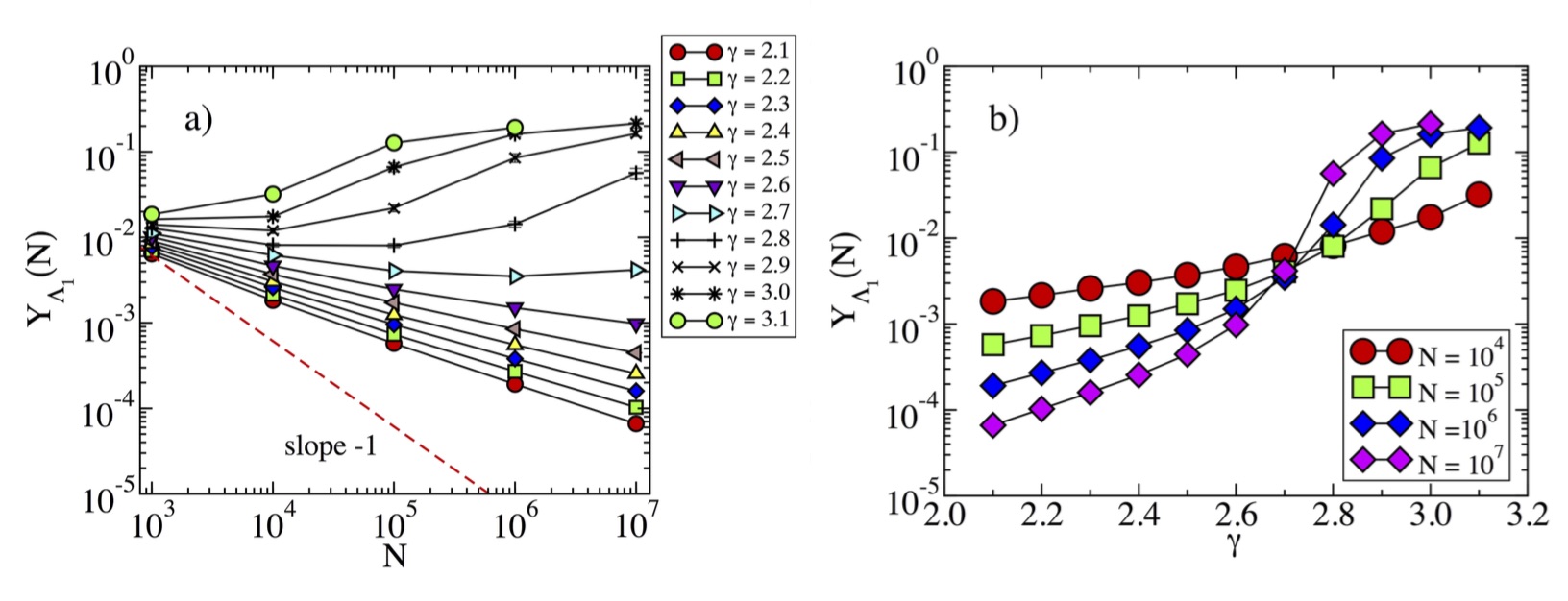} 
  \caption{(a) Inverse participation ratio as a function of the network
    size for the adjacency matrix of synthetic networks with different
    degree exponent $\gamma$.  For large $\gamma$, the IPR tends to
    saturate to a constant value for sufficiently large value of $N$.
    For $\gamma<5/2$, on the other hand, the behavior of the IPR can be
    fitted to $Y_{\LM}(N) \sim N^{-\alpha}$, with $\alpha < 1/2$.  The
    dashed line represents a power-law behavior $\sim N^{-1}$,
    corresponding to a delocalized IPR. (b) Inverse participation ratio
    as a function of the degree exponent $\gamma$ for different network
    sizes $N$. The plot confirms the presence of transition in the
    behavior of the IPR, located in the vicinity of
    $\gamma=5/2$. }
  \label{fig:Y4_vs}
\end{figure}

We study the localization properties of the PEV computed for synthetic
power-law distributed networks of growing size, generated using the
uncorrelated configuration model (UCM) \cite{ucmmodel}, a modification
of the standard configuration model~\cite{Molloy95,molloy98} designed to
avoid degree correlations \cite{mariancutofss}.  In order to explore the
presence or absence of localization, we analyze the scaling of
$Y_{\Lambda} (N)$
as function of $N$
as discussed above. In Fig.~\ref{fig:Y4_vs}(a) we apply this finite-size
scaling analysis to synthetic networks with different values of
$\gamma$.
In this and the following figures, statistical averages are performed
over at least 100 different network samples. Error bars are usually
smaller than the symbol sizes.  In the case of large $\gamma$
we observe an IPR tending to a constant for large $N$,
confirming the localization on the hubs predicted
by~\cite{Goltsev12,Nadakuditi13}.  The situation is however surprisingly
different for $\gamma<5/2$.
Thus, while according to Goltsev~\textit{et al.}~\cite{Goltsev12}, we
should expect a delocalized PEV and an IPR decreasing as $N^{-\alpha}$
with $\alpha = 1$,
we observe instead power-law decays with $N$,
with effective exponents $\alpha$
always smaller than $1/2$.
The change of behavior of the IPR can be further confirmed in
Figure~\ref{fig:Y4_vs}(b), where we plot the IPR as a function of the
degree exponent $\gamma$,
for different values of $N$.
While it is clear that for $\gamma \ge 2.7$
the IPR tends to a constant asymptotically, slow crossover effects do
not allow to draw firm conclusions based on numerics about the precise
value of $\gamma$
for which the behavior changes. However, since the dependence of the
largest eigenvalue on $N$
changes for $\gamma=5/2$~\cite{Chung03}
we expect the transition to take place exactly at $\gamma=5/2$:
simulation results are perfectly compatible with this result.

\begin{figure}[t]
  \centering
  \includegraphics[clip=true,trim=0cm 0cm 0cm 0cm,width=15cm]%
  {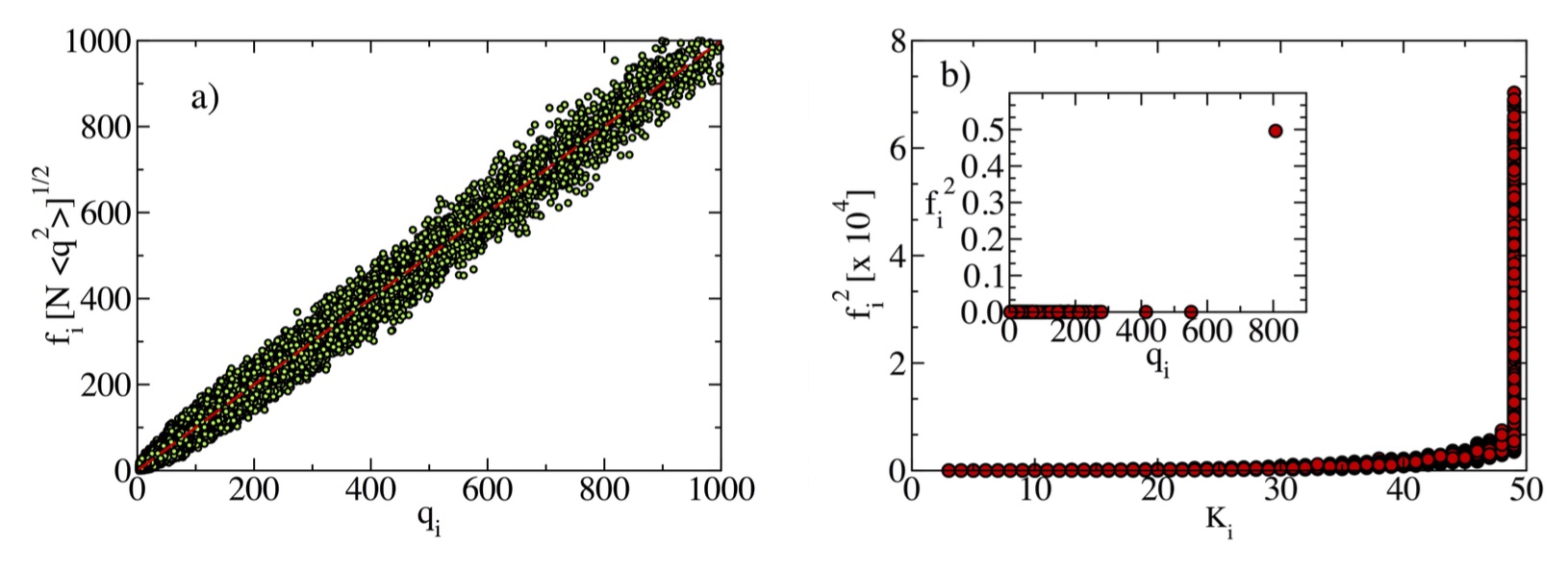} 
  \caption{(a) Rescaled scatter plot of $f_i [N \av{q^2}]^{1/2}$ as a
    function of $q_i$ for a synthetic network with $\gamma=2.1$ and size
    $N=10^6$. Data fits the expectation for the PEV in the annealed
    network approximation, Eq.~(\ref{eq:1}), with only small
    fluctuations.  (b, main) Scatter plot of the squared PEV components
    as a function of the $K$-core index for the adjacency matrix of a
    power-law synthetic network with $\gamma=2.1$ and size $N=10^6$.
    (b, inset) Scatter plot of the squared PEV components as a function
    of the degree $q_i$ in a synthetic network with $\gamma=3.5$ and
    size $N=10^6$.  }
  \label{fig:Y4_vs_2}
\end{figure}

The behavior at $\gamma<5/2$ can be understood mathematically by
observing that the largest eigenvalue in this regime, $\Lambda_1 = \AV$
\cite{Chung03}, coincides with the largest eigenvalue of the adjacency
matrix in the annealed network approximation.  The annealed network
approximation \cite{dorogovtsev07:_critic_phenom,Boguna09} consists in
replacing the actual, fixed, adjacency matrix by an average performed
over degree classes, taking the form
\begin{equation}
  \bar{a}_{ij} = \frac{q_j' P(q_i \vert q_j')}{N P(q_i)},
\end{equation}
where $ P(q \vert q')$ is the conditional probability that a link from a
node of degree $q'$ points to a node of degree $q$ \cite{alexei}. For
degree uncorrelated networks, with $ P(q \vert q') = q P(q)/\av{q}$
\cite{Dorogovtsev:2002}, we obtain an averaged adjacency matrix
\begin{equation}
  \bar{a}_{ij} = \frac{q_i q_j}{N \av{q}}.
  \label{eq:2}
\end{equation}
The matrix $\bar{a}_{ij}$
is semi-positive definite and therefore all its eigenvalues are
non-negative~\cite{Gantmacher}. Then considering that
$\mathrm{Tr}(\mathbf{\bar{a}}^2) = [\mathrm{Tr}(\mathbf{\bar{a}})]^2 =
(\av{q^2} / \av{q})^2$, where $\mathrm{Tr} (\cdot)$
is the trace operator, we have that $\bar{a}_{ij}$
has a unique non-zero eigenvalue
$\Lambda_\mathrm{an} = \av{q^2} / \av{q}$,
with associated principal eigenvector $f^\mathrm{an}_i \propto q_i$.
Applying the normalization condition $\sum_i f_i^2 = 1$,
we obtain the normalized form
\begin{equation}
  f_i^\mathrm{an} = \frac{q_i}{[N \av{q^2}]^{1/2}}.
\label{eq:1}
\end{equation}
Inserting the expression of $f_i^\mathrm{an}$ into Eq.~\eqref{eq:5}
yields
\begin{equation}
Y_{\LM}(N) \sim 1/N^{(3-\gamma)/2},\label{eq:3}
\end{equation}
that is, a decay with an exponent smaller than $1/2$, in agreement with
the results in Fig.~\ref{fig:Y4_vs}(b).  Fig.~\ref{fig:Y4_vs_2}(a)
confirms that also quenched synthetic networks have PEV components
proportional in average to the degree. Notice that
  Eq.~(\ref{eq:3}) is approximately true only in quenched networks for
  $\gamma<5/2$, since the condition leading to it, Eq.~(\ref{eq:1})
  fails at $\gamma>5/2$, see Fig.~\ref{fig:Y4_vs_2}(b,inset).

\begin{figure}[t]
  \centering
 \includegraphics[clip=true,trim=0cm 0cm 0cm 0cm,width=15cm]%
  {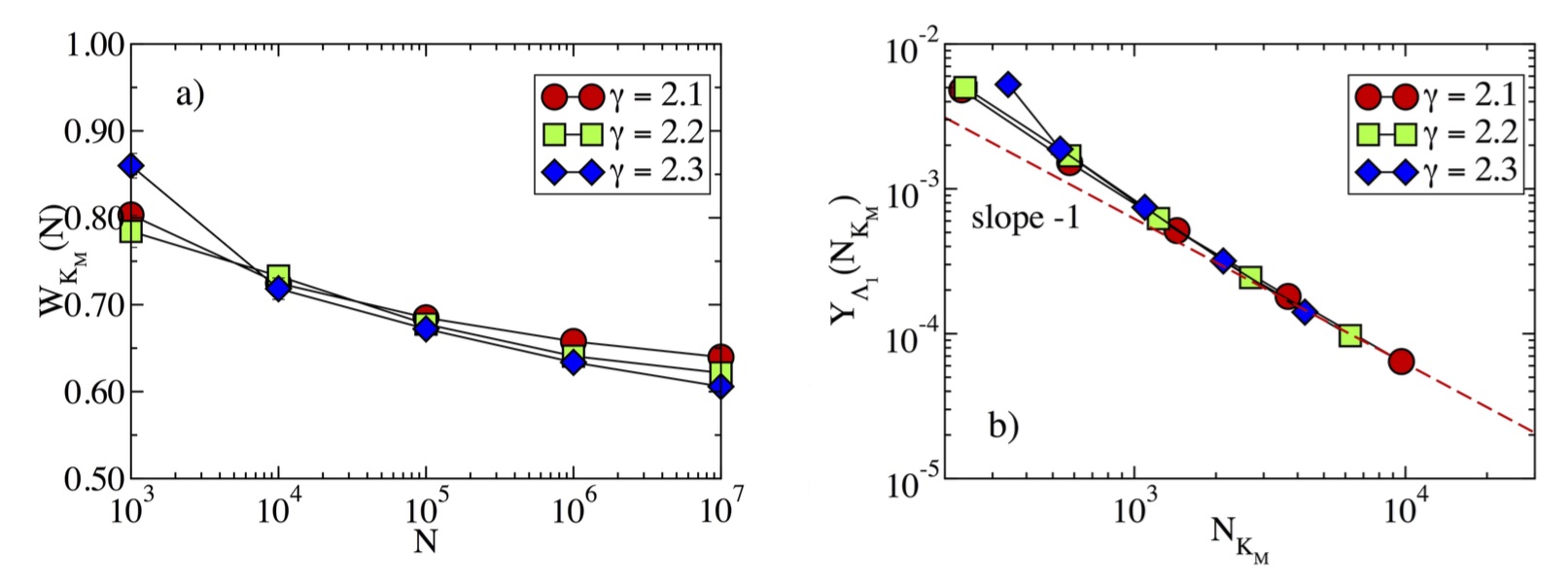} 
  \caption{(a) Total weight $W_{K_M}$ of the PEV on the nodes of the max
    $K$-core in synthetic networks as a function of size $N$. (b)
    Inverse participation ratio as a function of the size of the max
    $K$-core $N_{K_M}$.  The dashed line represents a power-law behavior
    $\sim N_{K_M}^{-1}$.  We can see the asymptotic behavior
    $Y_{\LM} \sim N_{K_M}^{-1}$, valid for large network sizes.  }
  \label{fig:Y4_vs_3}
\end{figure}

A more physical interpretation of the particular distribution of the PEV
in power-law networks with $\gamma<5/2$,
is that the PEV becomes \textit{effectively} localized on the max(imum)
$K$-core
of the network, defined as the set of nodes with the largest core index
$K_M$
in a $K$-core
decomposition~\cite{Seidman1983269,PhysRevLett.96.040601}.  The $K$-core
decomposition is an iterative procedure to classify vertices of a
network in layers of increasing density of connections. Starting with
the full graph, one removes the vertices with degree $q=1$,
i.e. with only one connection. This procedure is repeated until only
nodes with degree $q \geq 2$
are left. The removed nodes constitute the $K=1$-shell
and those remaining compose the $K=2$-core.
At the next step all vertices with degree $q = 2$
are removed, thus leaving the $K=3$-core.
The procedure is repeated iteratively. The maximum $K$-core
(of index $K_M$)
is the set of vertices such that one more iteration of the procedure
removes all of them. The line of argument leading to this interpretation
stems from combining the results of Ref.\cite{Goltsev12}, in which it is
proposed that, in epidemic spreading in complex networks
\cite{Pastor-Satorras2015}, infection activity is localized on the PEV,
with the observations in Ref.\cite{Castellano12}, in which the maximum
$K$-core
is identified as a subset of nodes sustaining epidemic activity for
$\gamma<5/2$.

We can see this effective localization on the maximum $K$-core in
different ways.  In the first place, in Fig.~\ref{fig:Y4_vs_2}(b,main)
we plot the squared components $f_i^2$ of the PEV for all vertices
against their corresponding $K$-core index. From this Figure we conclude
that all nodes with the largest $f_i$ components belong to the max
$K$-core.  The size of this max $K$-core, $N_{K_M}$, grows sublinearly
as a function of the network size as $N_{K_M} \sim N^{(3-\gamma)/2}$
~\cite{PhysRevLett.96.040601}.  However, despite this
sublinear growth, a finite fraction of the total PEV weight is
concentrated on this subset.
We check this fact in Fig.~\ref{fig:Y4_vs_3}(a): the total
weight of the nodes in the max $K$-core,
\begin{equation}
W_{K_M} = \sum_{i | i \in K_M} f_i^2,\label{eq:4}
\end{equation}
tends to a constant in the limit of large network
size, implying that more than half of the weight of the normalized
PEV resides over the max $K$-core.  Finally, the size dependence of the
max $K$-core translates, from Eq.~(\ref{eq:Methods_2}) in an IPR scaling
as $Y_{\LM} \sim {N_{K_M}}^{-1} \sim 1/N^{(3-\gamma)/2}$, in agreement
with the result obtained from the degree dependence of the PEV
components, $f_i \sim q_i$, see Eq.~(\ref{eq:3}). The relation between
IPR and max $K$-core size is satisfactorily checked in
Fig.~\ref{fig:Y4_vs_3}(b), where we observe it to be valid for large
network sizes.

\begin{figure}[t]
  \centering
  \includegraphics[clip=true,trim=0cm 0cm 0cm 2cm,width=7cm]%
  {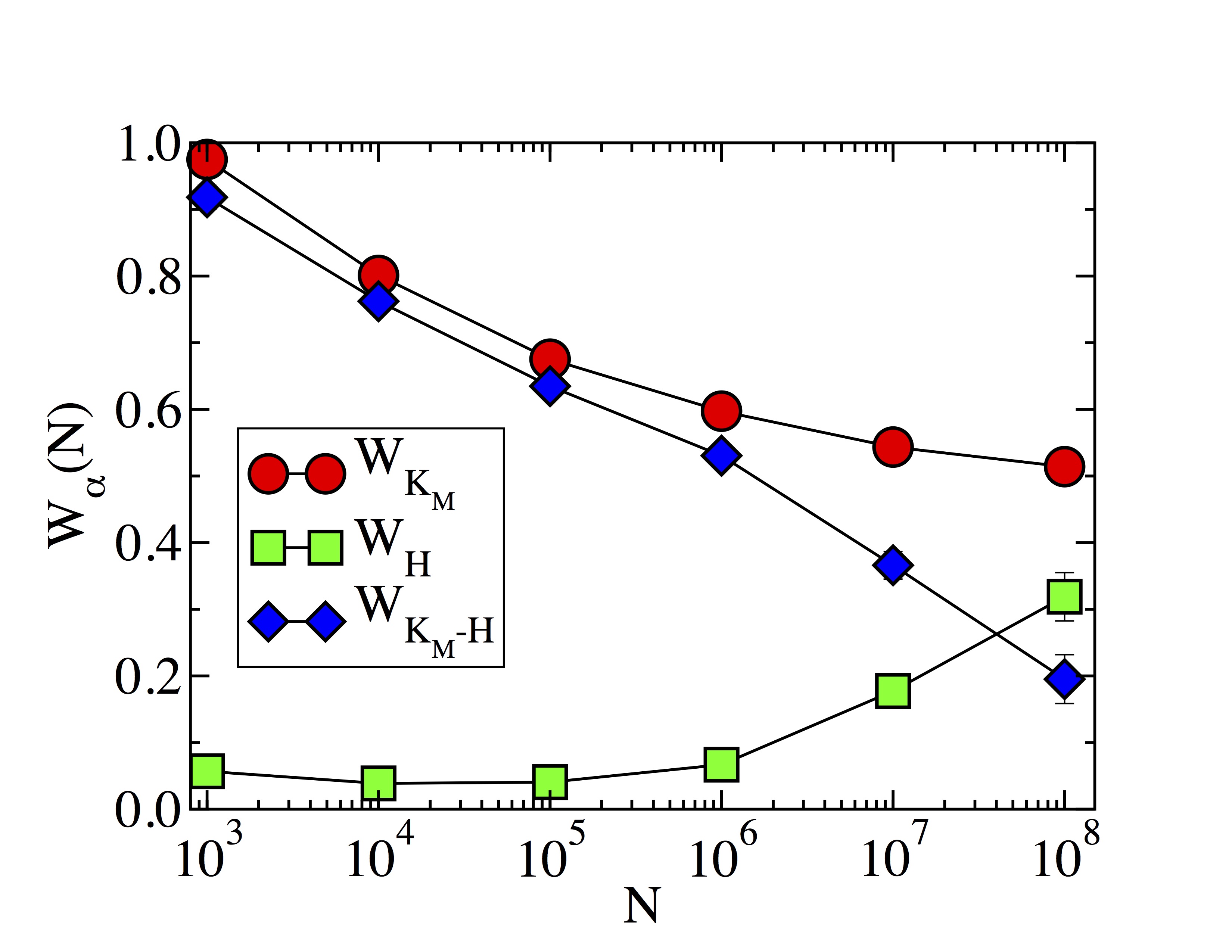} 
  \caption{Weight of the PEV as a function of the network size in
    power-law networks with degree exponent $\gamma=2.8$. The different
    functions correspond to: total weight of the nodes in max $K$-core,
    $W_{K_M}$; total weight in the hub, $W_H$; total weight in the max
    $K$-core, subtracting the hub, $W_{K_M-H}$.}
  \label{fig:partialW}
\end{figure}

For $\gamma>5/2$, instead, Figure~\ref{fig:Y4_vs_2}(b,inset) confirms
the localization of the PEV around the hub \cite{Goltsev12,Martin14},
displaying a disproportionately large component on the node with the
largest degree.  Notice that, irrespective of the value of $\gamma$,
with high probability the hub {\em belongs} to the max $K$-core.  What
changes in the two cases is that for $\gamma>5/2$ the hub alone carries
a finite fraction of the normalization weight ($f_i \sim
\mathcal{O}(1)$) while for $\gamma<5/2$ it carries a vanishing fraction,
and all nodes of the max $K$-core must be considered to have a finite
weight $W_{K_M}$. The behavior for $\gamma>3$ is clearly evident from
Fig.~\ref{fig:Y4_vs_2}(b,inset). In the case $5/2 < \gamma <3$, the
accumulation of a finite weight on the hub takes place for sufficiently
large $N$. This effect is observed in
Fig.~\ref{fig:partialW}, were we plot the total weight $W_{K_M}$ of the 
nodes in the max $K$-core, Eq.~(\ref{eq:4}), the total weight in the hub,
$W_H$, and the total weight in the max $K$-core, subtracting the hub,
$W_{K_M-H}$. As we can observe from this Figure, the weight at the hub
is small for network sizes $N<10^6$, but it then starts to increase, to
finally take over, for large network sizes $N> 10^7$. 

\subsection*{The non-backtracking centrality}

\begin{figure}[t]
  \centering
 \includegraphics[clip=true,trim=0cm 0cm 0cm 2cm,width=9cm]%
  {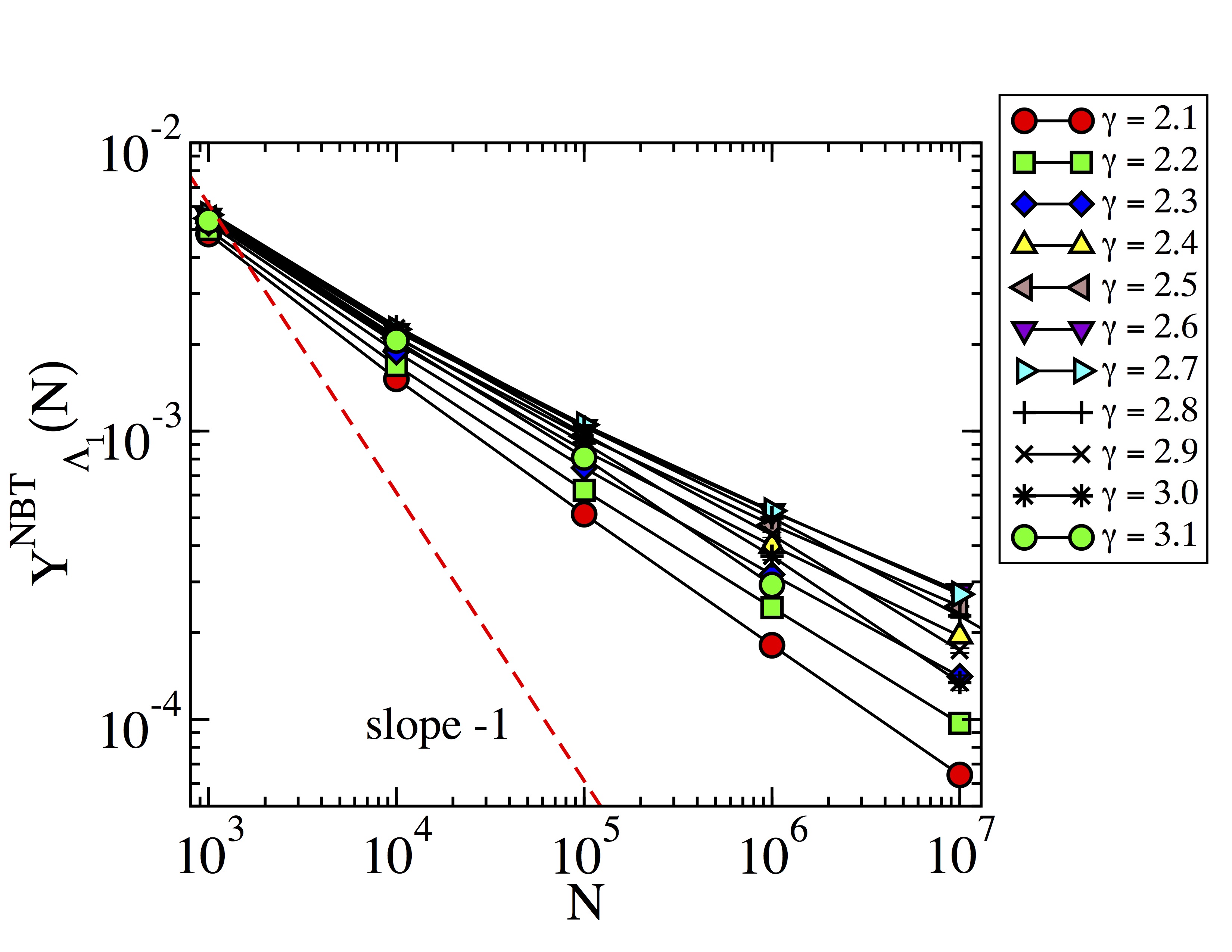}
  \caption{Inverse participation ratio as a function of the network size
    $N$ for the NBTC for power-law synthetic networks with different
    degree exponents $\gamma$.  The dashed line has slope $-1$
    indicating delocalization.}
  \label{fig:NBTPEVvsKs}

<\end{figure}

The observations presented here, together with the arguments provided by
Martin et \textit{al.}~\cite{Martin14}, hint that the EC is problematic
as a useful measure of centrality. For large values of $\gamma$, it is
affected by an exceedingly strong localization on the hub, arising as a
purely topological artifact: the hub is central because its neighbors
are central, but those in turn are central only because of the hub.  For
small values of $\gamma$, on the other hand, the observed relation
$f_i \sim q_i$ indicates that the eigenvector centrality provides
essentially the same information as the degree centrality. As an attempt
to correct the flaws of the EC, Martin et \textit{al.}~\cite{Martin14}
propose a modified centrality measure, the non-backtracking centrality
(NBTC), which is computed in terms of the non-backtracking matrix.  The
Hashimoto, or non-backtracking matrix (NBT)
\cite{hashimoto89,Krzakala24122013,Martin14}, is defined as follows: an
initially undirected network is converted into a directed one by
transforming each undirected edge into a pair of directed edges, each
pointing in opposite directions. If the initial undirected network has
$E$ edges, the NBT matrix is a $2E \times 2E$ matrix with rows and
columns corresponding to directed edges $i \to j$ with value
$B_{i \to j, l \to m} = \delta_{i,m} (1-\delta_{j,l})$, $\delta_{i,j}$
being the Kronecker symbol.  The components of the principal eigenvector
of the NBT matrix, $f_{i \to j}$ measure the centrality of vertex $i$
disregarding the contribution of vertex $j$. The NBT centrality of
vertex $j$ is given by the sum of these contributions for all neighbors
of $j$: $f_j^{NBT} = \sum_i A_{ij} f_{i \to j}$. The elements of the NBT
matrix count the number of non-backtracking walks in a graph and hence
remove self-feedback in the calculation of node centrality, thus
eliminating in principle the artificial topological enhancement of the
hub's centrality.

\begin{figure}[t]
  \centering
 \includegraphics[clip=true,trim=0cm 0cm 0cm 0cm,width=15cm]%
  {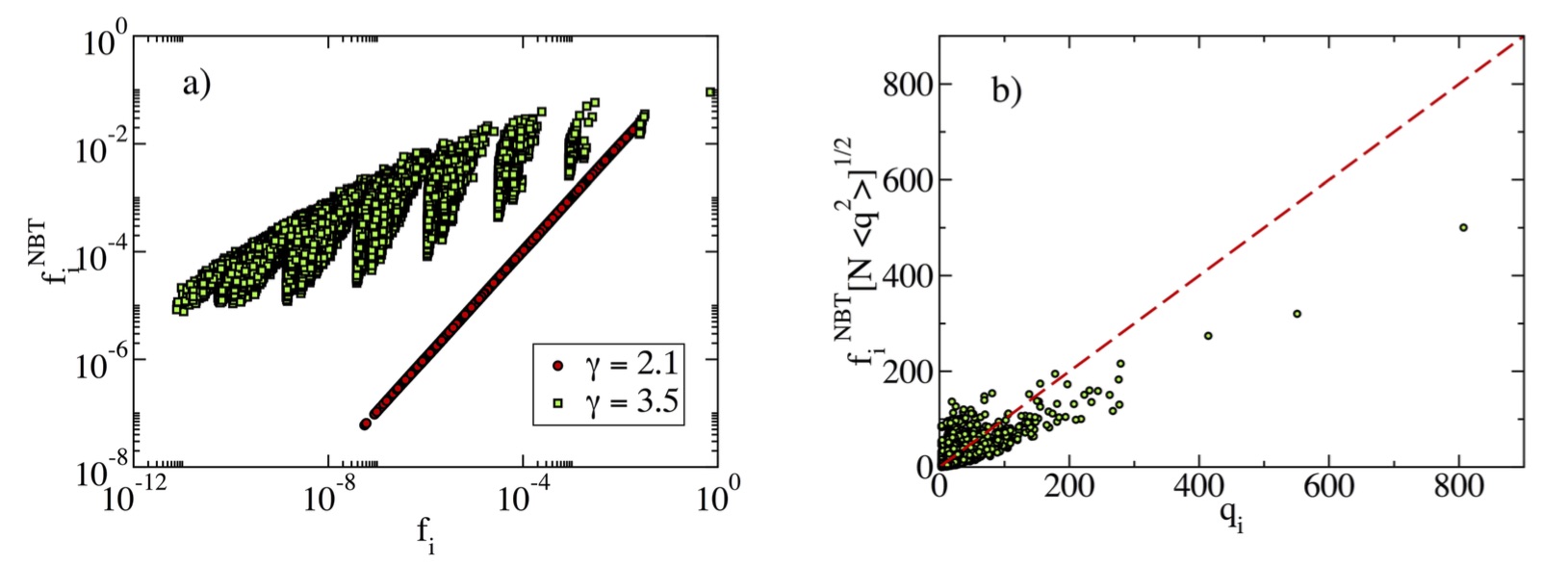}
  \caption{(a) Scatter plot of the NBTC centralities $f_i^\mathrm{NBT}$ as
    a function of the corresponding components of the PEV of the
    adjacency matrix $f_i$, in synthetic uncorrelated networks with a
    power-law degree distribution. Network size $N=10^6$. 
    (b) Rescaled scatter plot of the NBTC centralities
    $f_i^\mathrm{NBT} [N \av{q^2}]^{1/2}$ as a function of $q_i$ for a
    synthetic network with $\gamma=3.5$ and size $N=10^6$. }
  \label{fig:scatter_nbt_fi}
\end{figure}

\begin{figure*}[t]
  \centering
  \includegraphics[clip=true,trim=0cm 1cm 0cm 2cm,width=14cm]{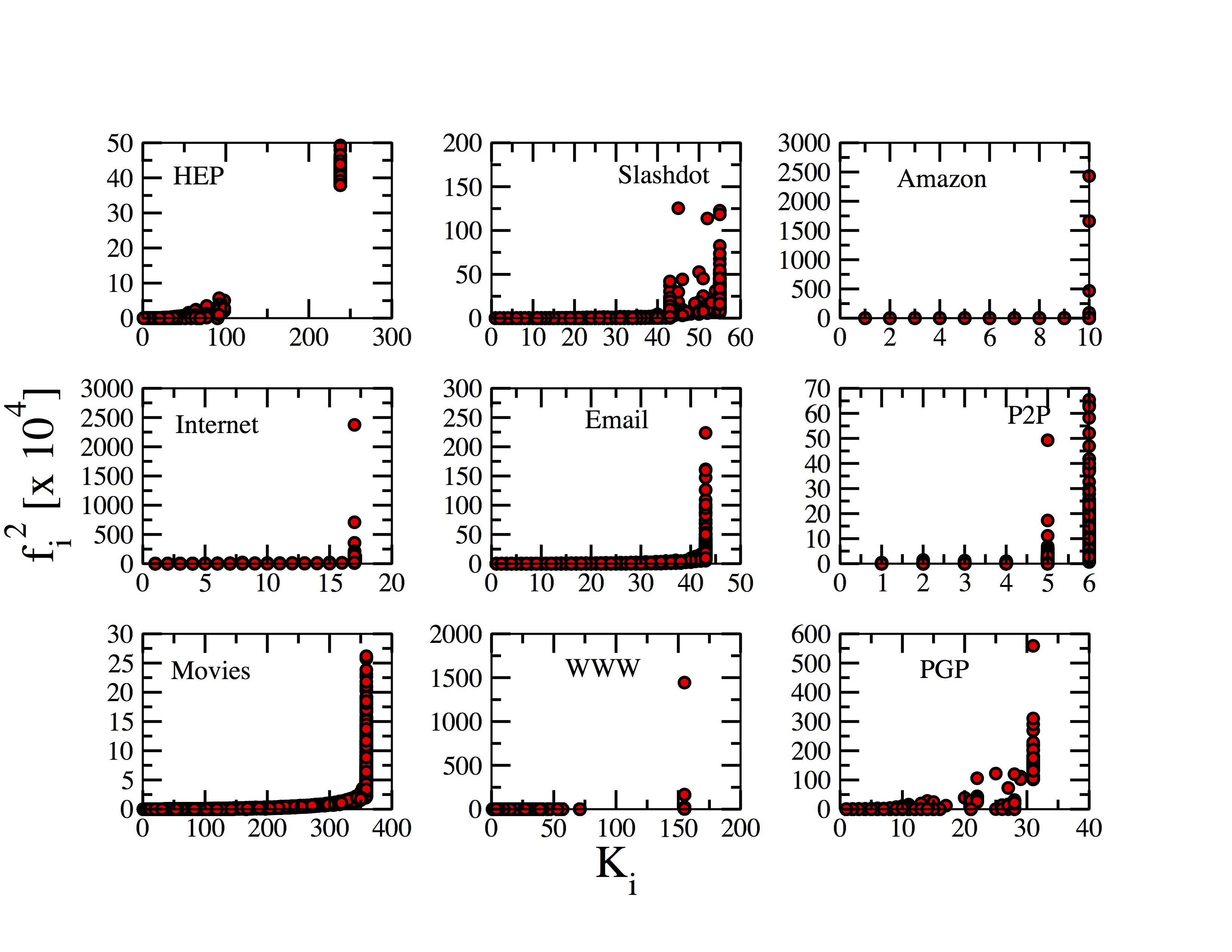}
  \caption{ Scatter plot of squared PEV components of the adjacency
    matrix of the real-world networks as a function of the $K$-core
    index.}
  \label{fig:Real1}

\end{figure*}

As Figure~\ref{fig:NBTPEVvsKs} shows, however, the NBTC is not free from
localization effects: For all values of $\gamma$ the NBTC is not
delocalized, i.e. $Y_{\LM}$ does not decrease as $1/N$ when increasing
$N$.  This fact can be understood for $\gamma<5/2$ in view of the
previous results.  The adjacency matrix PEV is localized on the max
$K$-core, which features many mutual interconnections: the centrality of
a node is only weakly affected by self-feedback, and removing the
contribution of backtracking paths has therefore little effect.  This is
confirmed by the scatter plot of the NBTC values $f_i^\mathrm{NBT}$ as a
function of the corresponding components $f_i$ of the adjacency matrix
PEV, computed for the same synthetic networks,
Figure~\ref{fig:scatter_nbt_fi}(a).  For $\gamma <5/2$ the two
quantities are very strongly correlated.  For $\gamma>5/2$ instead,
Figure~\ref{fig:scatter_nbt_fi}(a) shows that the NBT centrality is
truly different and uncorrelated from the adjacency matrix EC.  However,
as Fig.~\ref{fig:NBTPEVvsKs} shows, the NBT IPR, computed from the
components of the NBTC, decreases with the system size $N$ more slowly
than $N^{-1}$.  This is indicative that also in this case a localization
occurs on a mesoscopic subset, whose size grows sublinearly.
Figure~\ref{fig:scatter_nbt_fi}(b) shows that this localization is not
due to a strong correlation between the NBT centrality and the degree of
nodes, contrary to what happens for the EC for $\gamma<5/2$.

\subsection*{Eigenvector localization in real networks}
\label{sec:eigenv-local-real}

For real networks, which have fixed size and do not allow for a finite
size scaling analysis, localization is necessarily a more blurred
concept.  The value of $Y_{\LM}$ gauges how localized the PEV is,
but it does not permit to unambiguously declare a network localized or
not.  However, also in this case it is possible to detect, as in
synthetic networks, the existence of different localization modes.  We
consider here several real complex networks exhibiting large variations
in size, heterogeneity and degree correlations (see Methods and
Supplemental Material, SM, for details).

The linear relation between $f_i$ and the degree $q_i$ is not fulfilled
in real networks (see Supplementary Figure SF-1), probably due to the
presence of nontrivial degree correlations (see SM) 
which are absent in the synthetic networks. 
The effective localization on the max-$K$ core is
however still present in some cases.  In Fig.~\ref{fig:Real1} we plot for these
networks the squared PEV component $f_i^2$ as a function of the $K$-core
index.  In some cases (HEP, Movies) all nodes in the max $K$-core have a
comparable and large EC (as in synthetic networks for $\gamma<5/2$),
suggesting localization on the max $K$-core. In other cases (Internet,
Amazon) one or a few nodes have a disproportionately large value of
$f_i^2$, hinting at a localization around hubs, as in synthetic networks
for large $\gamma$.

To clarify the phenomenology we report in Table~\ref{table1} for each of
the real-world networks the values of the leading eigenvalue, and the
factors $\AV$ and $\sqrt{\km}$.  The analysis here is complicated
by the presence of degree correlations (see SM), which invalidate
the direct connection~\cite{Chung03} between $\Lambda_1$ and the largest
between $\sqrt{\km}$ and $\AV$~\cite{Goltsev12}.  However, in some cases
(Internet, Amazon)
the leading eigenvalue is much closer to $\sqrt{\km}$ than to $\AV$:
This suggests a localization around the hub and matches well with
Fig.~\ref{fig:Real1}.  In others the opposite is true: $\Lambda_1$ is
very far from $\sqrt{\km}$ and relatively close to $\AV$, hinting at a
localization on the max $K$-core, again in agreement with
Fig.~\ref{fig:Real1}.  In other cases (P2P, WWW), values are so close
that no conclusion can be drawn.

\begin{figure}[t]
  \centering
  \includegraphics[clip=true,trim=0cm 0cm 0cm 2cm,width=7cm]%
  {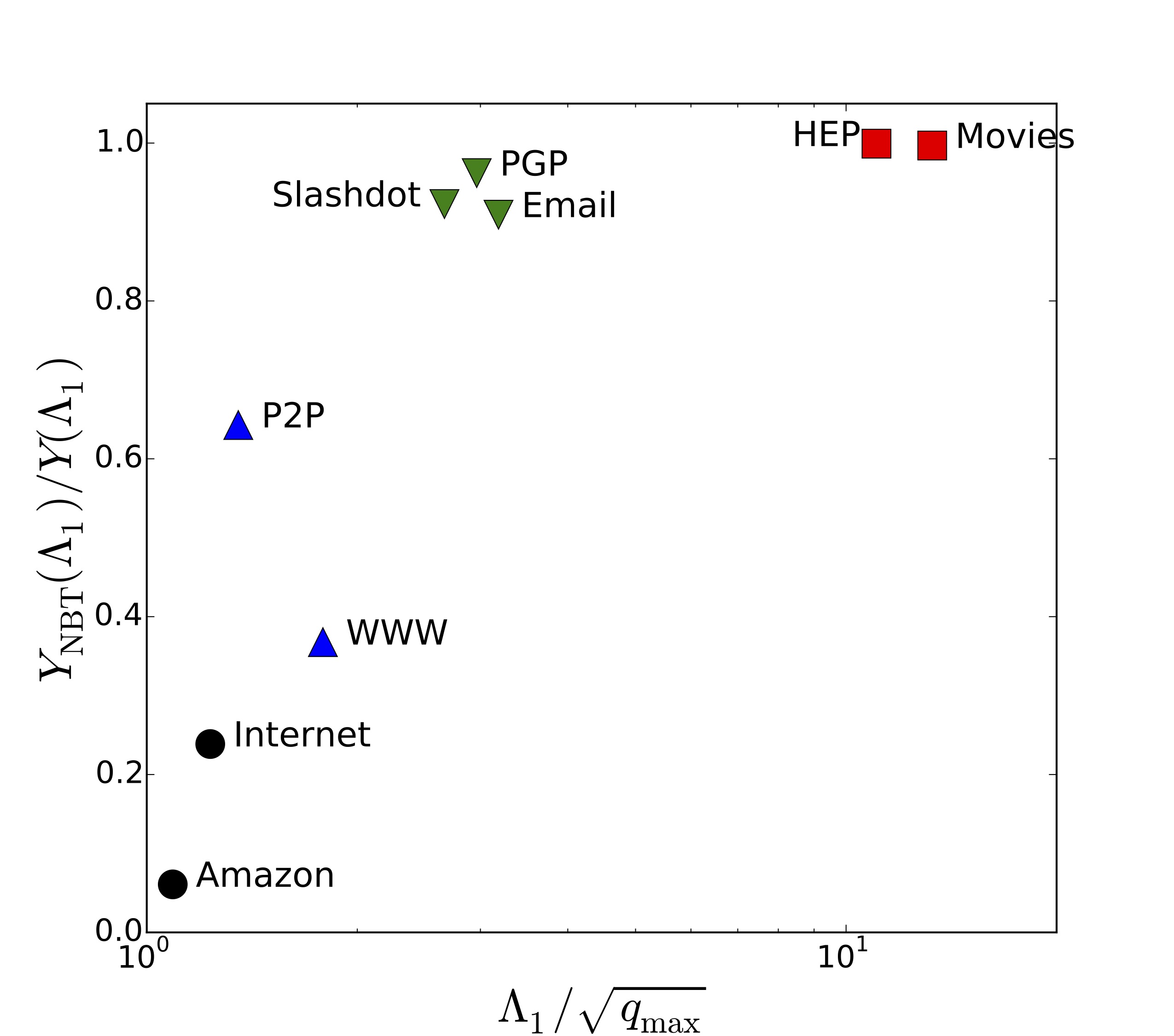} 
  \caption{Ratio between the NBTC IPR and the IPR of the adjacency
    matrix as a function of the ratio between the largest eigenvalue and
    the square root of the maximum degree, for the real networks
    considered. The symbol codes are: square for localization on the max
    $K$-core; circle for localization on the hub; triangle up for
    networks in which  $\sqrt{q_\mathrm{max}}$ is very close to
    $\av{q^2} / \av{q}$, so no conclusion can be drawn; triangle down
    for the rest of networks.}
  \label{fig:IPR_Ratio}
\end{figure}

A further confirmation of this picture is provided by the analysis of
the NBT centrality.  When localization occurs on hubs one expects the
elimination of backtracking paths to have a strong impact, as 
self-feedback effects are tamed.  In this case we expect the ratio between
the IPR for the NBTC and the IPR for the adjacency matrix to be
small.  On the contrary, when the localization occurs on the max
$K$-core, passing from the adjacency to the NBT matrix would not lead to
a big change and we expect the ratio to be close to 1.
Table~\ref{table1} confirms this expectation: the IPR ratio is small
when the leading eigenvalue $\Lambda_1$ is essentially given by
$\sqrt{\km}$ (localization on hubs) while it is close to 1 when
$\Lambda_1$ is closer to the $\AV$ factor (localization on the max
$K$-core). 
A visual representation of these results is provided in
Figure~\ref{fig:IPR_Ratio}, where we plot the IPR ratio as a function
of the ratio between $\Lambda_1$ and  $\sqrt{q_\mathrm{max}}$. As we can
see, networks in which the PEV is localized in the max $K$-core are
situated in the upper right corner of the panel, while the lower left
corner shows the networks with localization occurring on the hubs.

\section*{Discussion}

The properties of the principal eigenvector (PEV), and associated
largest eigenvalue, of the adjacency matrix defining a network have a
notable relevance as characterizing several features of its structure
and its effects on the behavior of dynamical processes running on top of
it.  Most important among these features is the role of the components of
the PEV as a measure of a node's importance, the so-called eigenvector
centrality. One of the properties of the PEV that has recently
attracted the interest of the statistical physics community is
its localization. In the case of networks with a power-law
degree distribution $P(q) \sim q^{-\gamma}$, initial research on this
subject \cite{Goltsev12,Martin14} suggested that, for $\gamma>5/2$, the
PEV is localized on the nodes with largest degree. On the other hand,
for $\gamma<5/2$, the PEV should be delocalized.

In this paper we have shown that eigenvector localization in
heterogeneous networks is described by a more complex picture. Thus, we
present evidence that for all power-law distributed networks the PEV 
is always localized to some extent. 
In the case of synthetic power-law distributed networks, we
observe that, while for mildly heterogeneous networks with $\gamma>5/2$
the PEV is indeed localized on the nodes with maximum degree (the hubs),
in the case of high heterogeneity, with $\gamma<5/2$, the PEV shows a
peculiar form of localization, its components $f_i$ being proportional
to the node's degree, $f_i \sim q_i$. This particular proportionality
induces an effective localization on the maximum $K$-core of the
network, defined as the core of maximum index in a $K$-core
decomposition. This max $K$-core concentrates a finite fraction of the
normalized weight of the PEV, despite the fact that the size of the max
$K$-core is sublinear with the network size. In the case of real world
networks, the elucidation of the PEV localization is not so clearcut. We
however provide evidence for an analogous scenario as that observed in
synthetic networks, where the nature of the localization of the PEV is
ruled by its associated largest eigenvalue $\Lambda_1$: When $\Lambda_1$
is close to the mean-field value $\AV$, localization on the max $K$-core
is expected. On the other hand, when the largest eigenvalue is close to
$\sqrt{\km}$, localization takes place on the hubs.

The results presented here give a new perspective on complex topologies
from several perspectives.  Firstly, it is common knowledge that
networks with $\gamma>3$ are fundamentally different from those with
$\gamma<3$ (scale-free networks) because the divergence of the second
moment of the degree distribution has a series of crucial effects. A
tacit corollary is that networks with $2< \gamma <3$ have essentially
the same properties.  Our paper, together with other recent 
results~\cite{Goltsev12}, points out that networks with exponent
$\gamma<5/2$ are in many respects qualitatively different from those
with $\gamma>5/2$.  Secondly, our results point out the weakness of
eigenvector centrality as a measure of centrality for power-law
networks. Indeed, for $\gamma<5/2$, eigenvector centrality does not
provide more information than degree centrality, while for $\gamma>5/2$
the eigenvector localization on the hubs arises as a purely topological
artifact. Alternative measures of centrality, based on the Hashimoto
non-backtracking matrix~\cite{Martin14,hashimoto89,Krzakala24122013} are
also not free from localization effects. Finally, from a dynamical point
of view, largest eigenvalues and the associated eigenvectors are
crucially related to the properties of processes on
networks~\cite{PhysRevE.71.036151,Goltsev12,2014arXiv1405.0483K} and
their localization effects should be taken properly into account when
developing theories relying on the structure of the adjacency matrix.

The localization properties described here call for a revision of our
present understanding of heterogeneous topologies.
Other networks properties, such as degree correlations, clustering or
the presence of a community structure, might play a role in the
localization of the PEV. The clarification of these effects, as well as
the understanding of the nature of the mesoscopic subgraph on which the
NBTC is localized for $\gamma > 5/2$, are still open questions, calling
for further scientific effort.

\section*{Methods}

\subsection*{Real networks analyzed}

We consider in our analysis the following  real networks datasets:

\begin{itemize}
\item \textbf{HEP}: Collaboration network between authors of papers
  submitted to the High Energy Physics section of the online preprint
  server arXiv. Each node is a scientist. Two scientists are connected
  by an edge if they have coauthored a preprint
  \cite{leskovec_snap_nets}. 

\item \textbf{Slashdot}: User network of the Slashdot technology news
  website. Nodes represent users, which can tag each other as friends or
  foes. An edge represents the presence of a tagging between two users
  \cite{leskovec_snap_2}. 

\item \textbf{Amazon}: Co-purchasing network from the online store
  Amazon. Nodes represent products, which are joined by edges if they are
  frequently purchased together \cite{Leskovec07}.

\item \textbf{Internet}: Internet map at the Autonomous System
  level, collected at the Oregon route
  server. Vertices represent autonomous systems (aggregations of
  Internet routers under the same administrative policy), while edges
  represent the existence of border gateway protocol (BGP) peer
  connections between the corresponding autonomous systems
  \cite{romuvespibook}. 

\item \textbf{Email}: Enron email communication network. Nodes represent
  email addresses. An edge joins two addresses if they have exchanged
  at least one email \cite{leskovec_snap_2}.

\item \textbf{P2P}: Gnutella peer-to-peer file sharing network. Nodes
  represent hosts in the Gnutella system. An edge stands for a
  connection between two Gnutella hosts \cite{leskovec_snap_nets}.

\item \textbf{Movies}: Network of movie actor collaborations obtained
  from the Internet Movie Database (IMDB). Each vertex represents an
  actor. Two actors are joined by an edge if they have co-starred at
  least one movie \cite{Barabasi:1999}.

\item \textbf{WWW}: Notre Dame web graph. Nodes represent web pages from
  University of Notre Dame. Edges indicate the presence of a hyperlink
  pointing from one page to another \cite{www99}.

\item \textbf{PGP}: Social network defined by the users of the
  pretty-good-privacy (PGP) encryption algorithm for secure information
  exchange. Vertices represent users of the PGP algorithm. An edge
  between two vertices indicates that each user has signed the
  encryption key of the other \cite{PhysRevE.70.056122}.

\end{itemize}

Some of this networks are actually directed. We have symmetrized them,
rendering them undirected, to perform our analyses.

\clearpage

\section*{Acknowledgments}

R.P.-S. acknowledges financial support from the Spanish MINECO, under
projects No. FIS2010-21781-C02-01 and FIS2013-47282-C2-2, EC
FET-Proactive Project MULTIPLEX (Grant No. 317532), and ICREA Academia,
funded by the Generalitat de Catalunya.

\section*{Author contributions statement}

R.P.-S. and C.C. designed the research. R.P.-S. performed the data
analysis. R.P.-S. and C.C wrote the paper.

\section*{Additional information}

\textbf{Competing financial interests} The authors declare no competing
financial interests. 

\clearpage

\begin{table*}
  \begin{center}
    \begin{tabular}{|c||c|c||c|c|c||c|c||c|}
      \hline
      Network  & $N$ & $\av{q}$ &$\AV$  & $\sqrt{q_{max}}$ & $\Lambda_1$ & $Y(\Lambda_1)$ & $Y(\Lambda^{NBT}_1)$ & IPR Ratio \\ \hline\hline
      HEP      & 12006 & 19.74 & 129.94 & 22.16            & 244.93      & 0.003890       & 0.003887             & 0.9993    \\ \hline 
      Slashdot & 82168 & 12.27 & 149.71 & 50.52            & 134.63      & 0.002174       & 0.002006             & 0.9228    \\ \hline 
      Amazon   & 403394 & 12.11 & 30.55  & 52.46            & 57.15       & 0.089122       & 0.005423             & 0.0608    \\ \hline
      Internet & 10790 & 4.16 & 259.46 & 48.34            & 59.58       & 0.066138       & 0.015783             & 0.2386    \\ \hline 
      Email    & 36692 & 10.02 & 140.08 & 37.19            & 118.42      & 0.003790       & 0.003446             & 0.9091    \\ \hline 
      P2P      & 62586 & 4.73 & 11.60  & 9.75             & 13.18       & 0.000921       & 0.000592             & 0.6429    \\ \hline 
      Movies   & 81860 & 89.53 & 594.92 & 61.55            & 817.36      & 0.000640       & 0.000638             & 0.9966    \\ \hline 
      WWW      & 325729 & 6.69 & 280.68 & 103.54           & 184.93      & 0.022726       & 0.008357             & 0.3677    \\ \hline 
      PGP      & 10680 & 4.55 & 18.88  & 14.32            & 42.44       & 0.016622       & 0.015989             & 0.9619    \\ \hline
    \end{tabular}
    \caption{Relevant metrics for the various real-world networks with
      and the measured value of the IPR ratio between
      $Y(\Lambda^{NBT}_1)$
      and $Y(\Lambda_1)$.
      Size and other information on the networks are provided in the
      Supplementary Information. }
    \label{table1}
  \end{center}
\end{table*}

\end{document}